\documentclass[12pt]{article}

\newcommand{\ma}{\ensuremath{\mathcal{A}}}
\newcommand{\me}{\ensuremath{\mathcal{E}}}
\newcommand{\mh}{\ensuremath{\mathcal{H}}}
\newcommand{\ms}{\ensuremath{\mathcal{S}}}
\newcommand{\msp}{\ensuremath{\mathcal{S}^{\prime}}}
\newcommand{\ha}{\ensuremath{\mathcal{H}_{\mathcal{A}}}}
\newcommand{\hs}{\ensuremath{\mathcal{H}_{\mathcal{S}}}}
\newcommand{\be}{\ensuremath{\begin{eqnarray}}}
\newcommand{\ee}{\ensuremath{\end{eqnarray}}}
\newcommand{\sj}{\ensuremath{|s_j>}}
\newcommand{\sk}{\ensuremath{|s_k>}}
\newcommand{\aj}{\ensuremath{|a_j>}}
\newcommand{\ak}{\ensuremath{|a_k>}}
\newcommand{\bsj}{\ensuremath{<s_j|}}

\begin{document}
\begin{center}
\Large{MEASUREMENTS AND DECOHERENCE} \\
\vspace{.2in} 
\large{Tulsi Dass}\\
\vspace{.15in}
\large{Chennai Mathematical Institute} \\ 
92, G.N. Chetty Road, T. Nagar, Chennai, 600017, India \\
Email : tulsi@cmi.ac.in
\end{center} 

\vspace{.18in}
\noindent
Lectures at the Nineth Frontier Meeting (organised by Km. MeeraMemorial 
Trust) at the Centre for Learning at Varadanahalli (near Bangalore), 
Jan 3-7, 2005.

\vspace{.18in}
\noindent
\textbf{Abstract}. A pedagogical and reasonably self-contained introduction 
to the measurement problems in quantum mechanics and their partial solution 
by environment-induced decoherence (plus some other important aspects of 
decoherence) is given.The point that decoherence does not solve the 
measurement problems completely is clearly brought out. The relevance of 
interpretation of quantum mechanics in this context is briefly discussed.

\newpage

\noindent
\textbf{Contents} 

\vspace{.12in}
\begin{enumerate} 
\item Introduction 
\item Density Operators and Reduced Density Operators
\item The Measurement Problem(s) in Quantum Mechanics \\
3.1 Measurements in traditional quantum mechanics; the problem of 
\hphantom{3.1 } macroscopic superpositions \\
3.2 Comparison with the classical case \\ 
3.3 The preferred basis problem
\item Decoherence \\
4.1 Basic concepts about decoherence \\
4.2 The Standard mechanism of environment-induced decoherence \\
4.3 Environment-induced decoherence in measurements \\
4.4 Pointer basis of the quantum apparatus \\
4.5 Environment-induced superselection rules \\
4.6 Decoherence in a soluble model; decoherence time-scale \\
4.7 Emergence of classicality
\item  Does Decoherence Completely Solve the Measurement Problems ?
\end{enumerate}

\newpage
\noindent
\textbf{I. Introduction}

\vspace{.12in}
During the past three and a half decades, there have been some important 
developments relating to the foundations of quantum mechanics (QM). The 
key development is the realisation that a realistic quantum system 
is not isolated; it is immersed in in an environment with which it 
continuously interacts. An important effect of this interaction is 
\emph{decoherence} --- the destruction of phase correlations in certain 
superpositions of  quantum states. Incorporation of this phenomenon in 
the quantum theoretic treatment of measurements serves to contribute 
substantially  towards the solution of the measurement problem which is 
essentially the problem of understanding as to how the superposition of 
quantum states of (the measured system + the apparatus) resulting from the 
measurement-interaction between the two results eventually into a unique 
pointer state. This development has implications relating to the problem of 
understanding the emergence of the classical world in a quantum universe 
and to deeper issues like the objective reality (of objects and phenomena). 

A large number of  articles (original papers and reviews) and books relating 
to various aspects of the `decoherence program' have already appeared in 
literature [1-13]. In these lectures, I shall briefly describe some 
important developments relating to this program.

\vspace{.2in}
\noindent
\textbf{II. Density Operators and Reduced Density Operators}

In this section we shall cover some elementary background relating to the 
density operators which will be used throughout the article. 

The \emph{trace} of an operator A in a Hilbert space \mh \ is defined as 
the sum of expectation values of A with respect to an orthonormal basis.:
\begin{eqnarray*}
Tr A = \sum_{i}<e_i|A|e_i>
\end{eqnarray*}
( the right hand side is well defined only for a subclass of operators-- the 
trace class operators); it is independent of the choice of the orthonormal 
basis.

A \emph{density operator} (also called \emph{density matrix} or 
\emph{statistical operator} in quantum mechanics literature) $\rho$ is a 
self adjoint, positive operator of unit trace :
\begin{eqnarray*}
\rho^{\dagger} = \rho; \hspace{.15in} \rho \geq 0; \hspace{.15in} 
Tr \rho = 1.
\end{eqnarray*}
[An operator A is \emph{positive} if $ <\psi|A|\psi> \geq 0 $ for all 
$\psi \in \mh$. Positivity, in fact, implies self-adjointness. The implied 
redundancy in the definition above, however, is harmless.] A convex 
combination (weighted sum) of density operators :
\begin{eqnarray}
\rho = \sum_i p_i \rho_i; \hspace{.15in} p_i \geq 0; \hspace{.15in} 
\sum_i p_i = 1 
\end{eqnarray}
is a density operator. 

Density operators in a quantum mechanical Hilbert space represent states. 
Those density operators which cannot be expressed as nontrivial convex 
combinations of other density operators [i.e. those not having more than one 
nonzero $p_i$ in Eq.(1)] represent pure states. These operators are of the 
form $ \rho = |\psi><\psi|$ (projection operators on one dimensional 
subspaces of \mh \ ) and satisfy 
the condition $\rho^2 = \rho$. Other (non-pure) states are called mixed 
states or mixtures. 

The expectation value of an observable ( self-adjoint operator) A in a state 
$\rho$ is givenby $Tr(\rho A)$. With $ \rho_i = |\psi_i><\psi_i|$ in Eq.(1), 
we have 
\begin{eqnarray}
Tr(\rho A) = \sum_i p_i Tr(\rho_i A) = \sum_i p_i <\psi_i| A |\psi_i>.
\end{eqnarray}
In Eq.(2), the averaging over the states $|\psi_i>$ reflects the irreducible 
probabilistic nature of quantum mechanics (QM); the second averaging with 
weights $p_i$ reflects the ignorance of the observer as treated in 
classical probability theory.

In the Schr$\ddot{o}$dinger picture, evolution of states is given in terms of 
the unitary evolution operators $U(t^{\prime},t)$ by
\begin{eqnarray}
\rho(t^{\prime}) = U(t^{\prime},t) \rho(t) U(t^{\prime},t)^{\dagger}.
\end{eqnarray}
For $ U(t^{\prime},t) = exp[-iH(t^{\prime}-t)/\hbar], $  it gives the von 
Neumann equation 
\begin{eqnarray}
i \hbar \frac{d\rho(t)}{dt} = [H, \rho(t)].
\end{eqnarray}

Given the Hilbert spaces $\mathcal{H}_1$ and $\mathcal{H}_2$  for two 
quantum mechanical systems $S_1$ and $S_2$ respectively, the Hilbert 
space for the combined system is the tensor product $\mh = \mathcal{H}_1 
\otimes \mathcal{H}_2$. Denoting by $I_1$ and $I_2$ the unit operators on 
$\mathcal{H}_1$ and $\mathcal{H}_2$, the operators A on $\mathcal{H}_1$ 
and  B  on $\mathcal{H}_2$ are represented as, respectively, the operators 
$A\otimes I_2$ and $I_1 \otimes B$ on $\mathcal{H}_1 \otimes \mathcal{H}_2$. 
Given orthonormal bases $ \{|e_i> \}$ in $\mathcal{H}_1$ and $ \{|f_r> \}$ 
in $\mathcal{H}_2$, the family $ \{ |e_i> \otimes |f_r> \} $ constitutes an 
orthonormal basis in $\mathcal{H}_1 \otimes \mathcal{H}_2$. 

Given an operator $A = A_1 \otimes A_2$ on $ \mathcal{H}_1 \otimes 
\mathcal{H}_2$, the \emph{partial traces} 
\begin{eqnarray*}
Tr_1(A) = (\sum_i<e_i|A_1|e_i>) A_2 , \hspace{.15in} 
Tr_2(A) = (\sum_r<f_r|A_2 |f_r>)A_1 
\end{eqnarray*}
are operators on $\mathcal{H}_2$ and $ \mathcal{H}_1$ respectively. These 
definitions are extended to general operators of $\mh_1 \otimes \mh_2$ 
(supposedly expressible as linear combinations of operators of the form 
$A_1 \otimes A_2$ ) by linearity. If $\rho$ is a density operator 
for the joint system on $\mathcal{H}_1 \otimes \mathcal{H}_2$, the operators 
\begin{eqnarray}
\rho_1 = Tr_2(\rho), \hspace{.2in} \rho_2 = Tr_1(\rho)
\end{eqnarray}
are referred to as the \emph{reduced density operators} for the systems 
$S_1$ and $S_2$ respectively; they incorporate the effect of the interaction 
between the two systems (as reflected in the expectation values of the 
observables of the two systems) on $S_1$ and $S_2$ respectively  in the 
following sense :
\begin{eqnarray*}
Tr [\rho(A \otimes I_2)] = Tr[\rho_1 A], \hspace{.15in} 
Tr [ \rho (I_1 \otimes B)] = Tr[ \rho_2 B].
\end{eqnarray*}
Replacing $\rho_i$ (i=1,2) and $\rho$ in Eq.(5) by $\rho_i(t)$ and 
$\rho(t)$, the von Neumann equation for $\rho(t)$ gives the effective 
evolution equations [ the so-called \emph{(generalized) master equations}] 
for $\rho_1(t)$ and $ \rho_2(t)$. In the presence of an interaction 
between $S_1$ and $S_2$, a unitary evolution of $\rho(t)$ generally leads to 
non-unitary evolutions for $ \rho_1(t)$ and $\rho_2(t)$. 

\noindent
Note (1). The decomposition [see Eq.(1)] of a mixed state into pure states 
is generally not unique. (There is nothing paradoxical or surprising about 
it.)\\
(2). Given a pure state on $\mathcal{H}_1 \otimes \mathcal{H}_2$ which is a 
superposition of the form
\begin{eqnarray}
|\psi> = \sum_{i,r} a_{ir} |e_i> \otimes |f_r>,
\end{eqnarray}
the corresponding reduced density operators $\rho_1$ and $\rho_2$ generally 
correspond to mixed states. For example 
\begin{eqnarray}
\rho_1 = Tr_2(|\psi><\psi|) = \sum_{s,i,j} a_{is} a_{js}^* |e_i><e_j|.
\end{eqnarray}
[Again, there is nothing paradoxical or surprising about it; the mixed state 
character of $\rho_1$ reflects the implicit ignorance of the precise (pure) 
state of $S_1$ when the joint system is given in the state (6).]\\
(3). On the Hilbert space \mh \ of a system S, a density operator of the form 
\begin{eqnarray}
\rho = \sum_i p_i |\psi_i><\psi_i|
\end{eqnarray}
represents an ensemble of systems (each of which is a copy of S) which 
randomly occupy (pure) states $|\psi_i>$, the probability of occupation of 
the state $\psi$ (i.e. the fraction of systems being in the state $\psi_i>$) 
being $p_i$. \\
(4). Since every self-adjoint operator can be diagonalized, a reduced density 
operator also admits a decomposition of the form (8); however, it generally 
does not admit the interpretation in the note (3) above. (See Ref.[31,32].)

\vspace{.2in}
\noindent
\textbf{III. The Measurement Problem(s) in Quantum Mechanics}

\vspace{.15in}
\noindent
\textbf{3.1 Measurements in traditional quantum mechanics; the problem of 
macroscopic superpositions [14,16-18,12,13]}

\vspace{.12in}
Let us consider the situation in which a quantum mechanical system \ms \ 
(the measured system; it is generally microscopic but need not always be so) 
with the associated Hilbert space \hs \ is given to be in a pure state 
$|\psi>$. We wish to perform a measurement of a physical quantity 
represented by the observable (self-adjoint operator) F on \hs \ . We assume, 
for simplicity, that F has a discrete non-degenerate spectrum : 
\begin{eqnarray}
F |s_j> = \lambda_j |s_j>.
\end{eqnarray} 
It is assumed that we can experimentally prepare any linear superposition of 
the states \sj \ . The state $|\psi>$ will generally be such a superposition: 
\begin{eqnarray}
|\psi> = \sum_j c_j \sj. 
\end{eqnarray} 

The measuring apparatus \ma \ is a macroscopic system chosen such that a 
suitable `pointer variable' associated with it takes values $ a_j$ in 
one-to-one correspondence with the $\lambda_js$. Different pointer positions 
are assumed to be macroscopically distinguishable. 

Niels Bohr had advocated the use of classical physics for the treatment of 
the apparatus. There 
is, however, no consistent formalism to describe the interaction between a 
quantum and a classical system. John von Neumann emphasized that quantum 
mechanics being (supposedly) a universally applicable theory, every system is 
basically quantum mechanical. To have a consistent theory of measurement, we 
must,therefore, treat the apparatus \ma \ 
quantum mechanically. Accordingly, we introduce a Hilbert space \ha \ for 
\ma \  and assume that the pointer positions $a_j$ are represented by  
the states \aj \ in this space. 

In the conventional treatment [14], one treats the combined system 
(\ms + \ma) as a closed system (ignoring the environment; we shall have to 
include it later) with the Hilbert space 
\begin{eqnarray}
\mathcal{H} = \hs \otimes \ha. 
\end{eqnarray}
For simplicity, we shall write $|\phi> |\chi>$ for the state 
$|\phi> \otimes |\chi>$. 
 
In his axiomatic scheme for QM, von Neumann postulated that 
there are two kinds of changes in quantum mechanical states : \\
(i) `Discontinuous, noncausal and instantaneous changes acting 
during experiments or measurements' (processes of the first kind); these he 
called  `arbitrary changes by measurement'. They are irreversible in  
nature. \\
(ii) `Continuous and cusal changes in the course of time' (processes of the 
second kind); these he called `automatic changes'. They are reversible. 

Changes of type (ii) are described by the traditional unitary evolution 
operators $U(t^{\prime},t) = exp[-iH(t^{\prime} - t)/\hbar].$ Those of 
type (i) are described, in the context of an observable F as above, by the 
replacement of the density operator $\rho$ of the system by 
\begin{eqnarray} 
 \rho^{\prime} = \sum_j P_j \rho P_j 
\end{eqnarray} 
where $ P_j = \sj \bsj$. We shall refer to this as the \emph{reduction 
process}. Under this process, the density operator  $\rho_{\psi} = 
|\psi><\psi|$ for the state (10) goes to 
\begin{eqnarray}
\rho_{\psi}^{\prime} = \sum_j P_j|\psi><\psi| P_j = \sum_j |c_j|^2 \sj \bsj 
\end{eqnarray} 
which is a mixture of the states $ P_j = \sj \bsj $ with weights $ w_j = 
|c_j|^2$. The density operator(13) represents an ensemble of systems (each a 
copy of \ms \ ) in which each system is in one of the states \sj \ , the 
fraction of the systems in the state \sj \ being $w_j$. 

von Neumann regarded the measuring process as consisting of two 
\mbox{stages:} \\
(i) the interaction between the measured system and the apparatus; this is 
governed by the unitary evolution process; \\
(ii) the act of observation; this involves the reduction process. 

Let $|a>$ be the 
initial ready state of the apparatus and $ U = U(t_f,t_i)$ the evolution 
operator for the interacting system (\ms + \ma) for the duration 
$ [t_i,t_f]$ of the measurement. If the initial state of the system is one 
of the \sj \ s, the pointer state after the measurement interaction is 
the corresponding \aj \ ; we describe this situation by the relation 
\begin{eqnarray}
U (\sj  |a>) = \sj  \aj.
\end{eqnarray} 
Writing the total Hamiltonianof the system-apparatus combine as $ H = 
H_{\ms} + H_{\ma}+ H_{int}$ , a standard way of satisfying the condition 
(14) is to employ the interaction of the von Neumann form 
\begin{eqnarray}
H_{int} = \sum_j \sj \bsj \otimes A_j 
\end{eqnarray}
where $A_j s$ are operators acting on \ha, \ and stipulate that , during 
the interval $[t_i,t_f]$ of the measurement interaction, the interaction 
term in the Hamiltonian dominates over the other two terms ( so that, 
effectively, $ H \simeq H_{int}$). This gives eq(14) with 
\begin{eqnarray}
\aj = e^{-iA_j (t_f -t_i)/\hbar} |a>. 
\end{eqnarray}
It must be emphasised, however, that Eq.(14) is a more general and clean way of 
describing the  effective interaction between the system and apparatus. The 
operator U acts essentially like the S-operator (traditionally called 
S-matrix) of quantum field theory [12].

Linearity of the evolution operator now implies that, when the initial 
state is (1), we must have 
\begin{eqnarray}
U(\sum_j c_j \sj  |a>) = \sum_j c_j \sj  \aj \equiv |\psi_f>. 
\end{eqnarray} 
Note that the right hand side of Eq.(17) is a superposition of the quantum 
states of the \emph{macroscopic} system (\ms + \ma). The operation 
represented by this equation is often referred to as \emph{premeasurement}.
This marks the completion of stage (i) of the measurement process. 

In the second stage, the reduction process becomes operative which transforms 
the density operator 
\begin{eqnarray}
|\psi_f><\psi_f| = \sum_{j,k}c_j^* c_k (\sk \bsj)(|a_k><a_j|) 
\end{eqnarray}
to 
\begin{eqnarray}
\rho^f_{\ms \ma} & = & \sum_j P_j |\psi_f><\psi_f| P_j \nonumber \\
                 & = & \sum_j |c_j|^2 (\sj \bsj)(\aj <a_j|). 
\end{eqnarray} 
This equation represents the joint state of system + apparatus at the 
completion of the measurement. It represents an ensemble of (system + 
apparatus systems) in which a fraction 
\begin{eqnarray}
p_j = |c_j|^2 
\end{eqnarray} 
appears in the j th product state in the summand. The final state of 
the apparatus is supposed to indicate the measured value of F. One concludes, 
therefore, that, given the system in the  state (10) and performing a 
measurement of the observable F, \\ 
(i) the measured values of the  observable F are the random numers 
$\lambda_j$ with probabilities $p_j$ given by eq(20); \\ 
(ii) when the measurement outcome is $\lambda_j$, the final state of the 
system is \sj.

The experimental verification of  the prediction (i) 
 consists in repeating this experiment a large numer of times 
[with  the same initial state (10)] and verifying (20) by invoking the 
frequency interpretation of probability. The prediction (ii) can be verified 
by repeating the experiment with the final state of the system in the previous 
experiment as the input state. Both predictions are in complete accord with 
experiment.

The main problem with the treatment of a quantum measurement given above is the 
ad hoc nature of the reduction [from Eq.(18) to (19)]. The so-called 
\emph{measurement 
problem} in quantum mechanics is essentially the problem of arriving at the 
above-mentioned random outcomes without introducing anything ad hoc in the 
theoretical 
treatment.This means that one should either give a convincing dynamical 
explanation of the reduction process or or else circumvent it. 

To the question : `where and how does the reduction process (19) take place ?' 
von Neumann's answer was that this is due to the involvement of human 
consciousness at the stage of actual observation of the outcome by the 
conscious observer. This proposal was later developed by London and Bauer 
[19] and Wigner [20]. This position, however, is not acceptable because the 
outcome of any experiment can be recorded by an appropriate device (say, a 
printer) and seen at convenience by a conscious observer. There is clearly 
no scope for the involvement of any subjective element in the explanation of 
the reduction process. 

Our description of the measurement problem will be very much incomplete 
without a mention of \textbf{Schr$\ddot{o}$dinger's cat}[21]. To emphasize 
the awkwardness of the macroscopic superpositions in Eq.(17), 
Schr$\ddot{o}$dinger introduced, in an experiment with two possible outcomes, 
a cat and a hypothetical device which would, in the event of one of the 
outcomes, kill 
the cat and leave it alive in the other case. One can now take the live and 
the dead states of the cat as the two pointer positions. The 
superpositions of Eq.(17) in this case would take the form
\begin{eqnarray}
c_1 |s_1> |LIVE \ CAT> + c_2 |s_2> |DEAD \ CAT>
\end{eqnarray}
etc. It has now become a tradition to refer to macroscopic superpositions 
as \emph{Schr$\ddot{o}$dinger cat states} (or simply \emph{cat states}).

\vspace{.15in}
\noindent
\textbf{3.2 Comparison with the classical case [32,17]}

\vspace{.12in}
It is instructive to compare the quantum mechanical situation described above 
with the classical case. Following Zurek [22], we adopt a Dirac-like 
notation $|\xi \}$ for pure states of a classical system S with phase space 
$\Gamma$ [$\xi$ = (q,p) is a point of $\Gamma$]. A general (pure or mixed) 
state of S is represented by a density function $\rho (\xi)$ on the phase 
space; pure states correspond to the $\delta$-function densities 
$\rho_{\xi}(\xi^{\prime}) = \delta (\xi, \xi^{\prime}).$

In classical mechanics (CM), there is no analogue of the superposition of 
pure states in quantum mechanics. The analogue of Eq.(1) of QM in CM is 
\begin{eqnarray}
\rho (\xi) = \sum_ip_i \rho_i (\xi).
\end{eqnarray}
(A convex comination of densities is a density.) The expansion of a 
mixed state in terms of pure states, however, is of the form
\begin{eqnarray}
\rho(\xi) = \int \rho(\xi^{\prime}) \rho_{\xi}(\xi^{\prime}) d\xi^{\prime}.
\end{eqnarray}
We shall formally consider (23) as a special case of (22) (with appropriate 
understanding about the `summations').

Clearly, classical pure states correspond, not to vectors in Hilbert space, 
but to the projection operators:
\begin{eqnarray}
|. \} \leftrightarrow |.><.|.
\end{eqnarray}

Given two systems $S_1$ and $S_2$ with phase spaces $\Gamma_1$ and $\Gamma_2$, 
a pure state of the combined system [ a point in the phase space $\Gamma_1 
\times \Gamma_2$ (Catesian product)] may be represented as $|\zeta \} = 
|\xi^{(1)} \} |\xi^{(2)} \}$ with $ \xi^{(i)} \in \Gamma_i $ (i=1,2). The time 
development of a state $ |\zeta(0) \} = |\xi^{(1)}(0) \} |\xi^{(2)}(0) \}$ 
gives the unique state $|\zeta(t) \} = |\xi^{(1)}(t)\}|\xi^{(2)}(t)\}$ 
where $\xi^{(i)}(t)$ is the phase space trajectory of the initial point 
$\xi^{(i)}(0)$ under the Hamiltonian evolution of the system $S_i$ (which 
may involve interaction between the systems $S_1$ and $S_2$). Both classical 
and quantum evolutions, therefore, preserve purity of states. 

In a measurement situation with $S_1 = \mathcal{S}$ and $S_2 = \ma$(now both 
considered as classical systems), we represent the (pure) states of the 
combined system as $|\zeta\} = |s\} |a\}$. As in the quantum case, there is 
a one-to-one correspondence between the states $|s_i\}$ of the system and 
the pointer states $|a_i\}$ of the apparatus. The classical analogue of Eq.(14) is (denoting the `ready' state of the pointer by $|a_0\}$)
\begin{eqnarray}
|s_j\} |a_0\} \longrightarrow |s_j\} |a_j\}
\end{eqnarray}
where the arrow represents the classical evolution of the combined system 
during the interval $[t_i,t_f]$ of the measurement interaction. There is, 
however, no classical analogue of Eq.(17); instead, we have 
\begin{eqnarray}
(\sum_j p_j |s_j\}) |a_0 \} \longrightarrow \sum_j p_j |s_j\} |a_j\}.
\end{eqnarray}
The right hand side has an obvious interpretation. There is no problem in 
classical mechanics analogous to the measurement problem in QM described 
above.

In the limit in which the left hand side of Eq.(26) goes to that of Eq.(25), 
the right hand side of Eq.(26) also goes to that of Eq.(25). Zurek refers 
to this as the complete information limit.

\vspace{.15in}
\noindent
\textbf{3.3 The preferred basis problem}

\vspace{.12in}
Apart from the problem of macroscopic superpositions, there is  another 
problem in  quantum measurement theory : the \emph{preferred 
basis problem}. This arises because the expansion in the final pre-measurement 
system-apparatus state $ |\psi_f>$ in Eq.(17) is generally not unique. If 
this state admits another expansion :
\begin{eqnarray}
|\psi_f> = \sum_k c_k^{\prime} |s_k^{\prime}> |a_k^{\prime}> ,
\end{eqnarray}
then the same state $|\psi_f>$ could correspond to measurement of any of the 
two observables 
\begin{eqnarray} 
F = \sum_j \lambda_j \sj \bsj ; \hspace{.2in} 
F^{\prime} = \sum_k \lambda_k^{\prime} |s_k^{\prime}><s_k^{\prime}| 
\end{eqnarray}
which may, in general, be even non-commuting as the example below shows. If 
the experimenter planned for a measurement of F, she/he cannot be sure that 
the pointer indicates a value of F  (and not that of $F^{\prime}$).

\vspace{.12in}
\noindent 
Example. We take \ms \ and \ma \ to be two-state systems; the two Hilbert 
spaces  are, therefore, two dimensional. Let 
\begin{displaymath} 
F = \sigma_z =  \left( \begin{array}{cc}
 1  & 0 \\
 0  & -1 
 \end{array} \right) \hspace{.2in}
 F^{\prime} = \sigma_x = \left( \begin{array}{cc} 
 0  &  1 \\
 1  &  0 
 \end{array} \right)
 \end{displaymath} 
 and let $ |z, \pm>, |x, \pm>$ be their respective eigenstates. Then we have, 
 for example,
 \begin{eqnarray}
 |\psi_f> & \equiv & \frac{1}{\sqrt{2}} [|z, +>_{\ms} |z, ->_{\ma} - 
 |z, ->_{\ms} |z,+>_{\ma}]  \nonumber \\
 & = & \frac{1}{\sqrt{2}}[ |x,+>_{\ms} |x, ->_{\ma} - 
 |x,->_{\ms} |x, +>_{\ma}]. 
\end{eqnarray}

\vspace{.2in}
\noindent
\textbf{IV. Decoherence}

\vspace{.12in}
In this section, we shall present a brief but reasonably self contained 
account of some important aspects of decoherence. 

\vspace{.15in}
\noindent
\textbf{4.1 Basic concepts about decoherence} 

\vspace{.12in}
Given a quantum mechanical state of some system $\mathcal{S}^{\prime}$  
(we shall reserve the symbols \ms \ and \ma \ to denote the measured system 
and appartus in a measurement context) which is a superposition of the form 
\begin{eqnarray}
|\phi> = \sum_j a_j |\phi_j> 
\end{eqnarray}
the corresponding density operator is 
\begin{eqnarray}
\rho_{\phi} = \sum_{j,k}a_j a_k^* |\phi_j><\phi_k|.
\end{eqnarray}
The off-diagonal terms in eq(31) contain phase correlations between 
different terms in the superposition (30)(phases of the complex amplitudes 
$a_j$ appear only in these terms). Quite often, one is interested in 
getting rid of these terms [as, for example, was the situation in the 
previous section with $\mathcal{S}^{\prime} = \ms + \ma$ and 
$ |\phi> = |\psi_f>$ of eq(17)]. \emph{Decoherence} is the general term 
employed to represent any phenomenon/process involving the disappearance 
(or strong suppression) of such off-diagonal terms. 

When the system in question is a macroscopic system (for example, a 
measurement apparatus), a common occurrence is the \emph{environment-induced  
decoherence} which is caused by the interaction of the system with the 
environment in which it is located. This environment consists of two parts: 
(i) the external environment consisting of air molecules, photons etc around 
the system, and (ii) the internal environment which is explained below.

In theoretical work, a macroscopic system is generally described in 
terms of the so-called \emph{collective observables} [12] like the position 
of a pointer, angular position of a pendulum, etc. (The word `collective' 
refers to the fact that these observables represent some property of the 
system as a whole 
rather than that of a few microscopic constituents.) The system, in its 
microscopic structure, consists of a large number of particles. Its fundamental 
observables are the canonical coordinate-momentum pairs, spin operators 
and operators for internal properties (like charge, isospin etc) for the 
constituent particles. Replacing these by an equivalent (in the sense that 
they generate the same algebra of observables) set of observables which 
includes the collective observables (and their conjugate partners) as a 
subset, then the observables in this set other than the collective obervables 
represent a quantum system called the internal environment. The full quantum 
mechanical system representing the totality of the constuents of the 
macroscopic system is thus formally divided into a \emph{collective system} 
(represented by 
the collective observables and their conjugate partners) and the 
\emph{(internal) environment}. 

The (system + environment ) combine is to be treated as a closed system. 
Since, strictly speaking, only the universe as a whole is a closed system,  
the external environment for a system should be the `rest of the universe'. 
One would like to have a suitable criterion for a smaller subsystem of this 
to be the effective external environment. A simple criterion is that, if 
one is considering a phenomen involving a time interval of order T 
(for example, in a Stern- Gerlach experiment, T is of the order of the 
time  taken by a typical atom in the beam  from the source to the detector), 
then the particles (including photons) in a sphere of radius of the order 
of cT (where c is the velocity of light) around the system in 
question should be included in the 
external environment. The effect of including interaction with the environment 
in the dynamics of the system depends (apart from the nature  of the 
interaction) essentially on the fact that the number of degrees of freedom 
in the environment is large. Some ambiguity in the identification of the 
environment is, therefore, of no consequence. 

Inclusion of interaction with the environment in the quantum theoretic 
treatment of a system results in, besides decoherence effects, energy 
transfer between the system and the environment which generally leads to 
dissipative effects in the dynamics of the system; at macroscopic level, 
they manifest themselves as friction, viscosity etc. In a quantum theoretic 
treatment of macroscopic systems, therefore, decoherence and dissipation 
effects appear side-by-side. 

\vspace{.15in}
\noindent
\textbf{4.2 Standard mechanism of environment-induced decoherence}

\vspace{.12in}
The Hilbert space for the (system $\mathcal{S}^{\prime}$ plus the 
environment \me \  ) is the tensor product $ \mh = \mh_{\msp}\otimes 
\mh_{\me}$. The total hamiltonian for the combine may be written as (in 
obvious notation)
\begin{eqnarray}
H = H_{\msp} \otimes I_{\me} + I_{\msp}\otimes H_{\me} +  H_{int}.
\end{eqnarray}
Dynamics of the combined system is given by the von Neumann equation (4). 
Effects of interaction with the environment on the dynamics of the system 
are described by the master equation for the reduced density operator 
$ \rho_{\msp} = Tr_{\me}(\rho)$. 

We shall assume $H_{int}$ to be of the von Neumann form [24] [see Eq.(15)]
\begin{eqnarray}
H_{int} = \sum _n |n><n| \otimes B_n.
\end{eqnarray}
As in the treatment of measurements in the previous section, we shall assume 
that  $H_{int}$ dominates over the other two terms in Eq.(32). (As we shall 
see below, decoherence is a very fast process; effects of the first two terms 
during the extremely short time intervals involved are relatively 
insignificant.) Assuming the initial state of the combined system as 
$(\sum_nb_n|n>)|\chi_0>$ and proceeding as in the previous section, we have 
the following analogues of Eqs.(14), (17) and (18):
\begin{eqnarray}
|n> |\chi_0> \rightarrow e^{-iH_{int}t/\hbar} |n> |\chi_0> = 
|n> |\chi_n(t)>  \hspace{.6in} \\
|\phi(0)> \equiv (\sum_n b_n |n>)|\chi_0> \rightarrow \sum_n b_n |n> 
|\chi_n(t)> \equiv |\phi(t)> \\
\rho(t) \equiv |\phi(t)><\phi(t)| = \sum_{m,n}b_n b_m^* (|n><m|) \otimes 
(|\chi_n(t)><\chi_m(t)|).
\end{eqnarray}
The reduced density operator for \msp \ corresponding to the density operator 
of Eq.(36) is 
\begin{eqnarray}
\rho_{\msp}(t) = Tr_{\me}\rho(t) = \sum_{m,n} b_n b_m^* |n><m| 
<\chi_m(t)|\chi_n(t)>.
\end{eqnarray}
Studies with fairly realistic models of the environment \me \ show that, for t 
large compared to the decoherence time scale $\tau_d$ [which is usually very 
small; see, for example, Eq.(85)], 
\begin{eqnarray}
<\chi_m(t)|\chi_n(t)> \longrightarrow \delta_{mn}.
\end{eqnarray}
This implies 
\begin{eqnarray}
\rho_{\msp}(t) \longrightarrow \sum_n|b_n|^2 |n><n|
\end{eqnarray}
which has no interference terms between the different $|n>$ states.

As a concrete example [3,11], let \msp \ be a two-state system with a basis 
$|u>, |d>$ (u for `up', d for `down') for $\mh_{\msp}$ and \me \ a system 
consisting of N two-state systems (N large) with basis $ |u_k>,|d_k>$ (k = 
1,...,N) for $\mh_{\me}$. We take 
\begin{eqnarray}
H \simeq  H_{int} = (|u><u| - |d><d|)\otimes \hspace{1.5in} \nonumber \\
\left[ \sum_{k=1}^N g_k 
(|u_k><u_k| - |d_k><d_k|)\otimes _{k^{\prime}\neq k} I_{k^{\prime}} \right].
\end{eqnarray}
The initial state 
\begin{eqnarray}
|\phi(0)> = (a|u> +b|d>) \otimes_{k=1}^N (\alpha_k |u_k> + \beta_k |d_k>)
\end{eqnarray}
evolves to 
\begin{eqnarray}
|\phi(t)> = a |u> |e_u(t)> + b |d> |e_d(t)>
\end{eqnarray}
where
\begin{eqnarray}
|e_u(t)> = |e_d(-t)> = \otimes_{k=1}^N (\alpha_k e^{ig_kt}|u_k> + 
\beta_ke^{-ig_kt}|d_k>).
\end{eqnarray}
The relevant reduced density operator is 
\begin{eqnarray}
\rho_{\msp}(t)& =& Tr_{\me} (|\phi(t)><\phi(t)|) \hspace{1.5in} \nonumber \\
&=& |a|^2 |u><u| + |b|^2 |d><d| + \nonumber \\
z(t) ab^* |u><d| + z^*(t)a^*b |d><u|
\end{eqnarray}
where 
\begin{eqnarray}
z(t) = <e_u(t)|e_d(t)> = \Pi_{k=1}^N (|\alpha_k|^2 e^{ig_kt} + 
|\beta_k|^2e^{-ig_kt}).
\end{eqnarray}
This gives 
\begin{eqnarray}
|z(t)|^2 = \Pi_{k=1}^N \{ 1 + [(|\alpha_k|^2 -|\beta_k|^2)^2 -1]sin^22g_kt \}.
\end{eqnarray}
We have z(0) =1. For those initial environment states which make the square 
bracket in Eq.(46) vanish for each k, z(t) =1 for all t and the interference 
terms between $|u>$ and $|d>$ in Eq.(44) remain present for all times. For 
generic environments, however, these terms are generally nonzero.Note that 
\begin{eqnarray}
 |z(t)| = |<e_u(t)|e_d(t)>| \leq 1.  \hspace{1.5in} \\
 <z(t)> \equiv lim_{T\rightarrow \infty}T^{-1}\int_{t-T/2}^{t+T/2}
z(t^{\prime})d t^{\prime} = 0. \hspace{1.1in} \\
 <|z(t)|^2>  =  2^{-N} \Pi_{k=1}^N [1 + 
(|\alpha_k|^2 - |\beta_k|^2)^2] 
 \rightarrow  0 \hspace{.12in} as \hspace{.12in} N  \rightarrow \infty.
\end{eqnarray}
This shows that, for large generic environments, the interference terms in 
Eq.(44) are strongly suppressed.

\vspace{.15in}
\noindent
\textbf{4.3 Environment induced decoherence in measurements}

\vspace{.12in}
We now include interaction with the environment \me \ in the quantum 
mechanical treatment of  measurements  given in the previous section. 
Taking \msp\  = \ms \ + \ma \ in the treatment of section 4.2, we have 
\begin{eqnarray}
\mh = \hs \otimes \ha \otimes \mh_{\me}.
\end{eqnarray}

Let the initial state of the environment be $|e>$. We assume the initial 
state of the combined system ($\ms + \ma + \me $) to be 
\begin{eqnarray}
|\phi(0)> = (\sum_jc_j \sj )|a> |e>.
\end{eqnarray}
The measurement interaction between the system and the apparatus takes it to 
the state [see Eq.(17)]
\begin{eqnarray}
(\sum_j c_j \sj \aj ) |e>.
\end{eqnarray}
Interaction of the system $\msp = \ms + \ma $ with the environment, acting as in section 4.2, takes the state (52 to [see Eq.(35)]
\begin{eqnarray}
|\phi_f> = \sum_j c_j \sj \aj |e_j>.
\end{eqnarray}
This gives the density operator
\begin{eqnarray}
\rho^f_{\ms \ma \me} = |\phi_f><\phi_f| = \sum_{j,k}c_j^*c_k(\sk \bsj)
(\ak <a_j|)(|e_k><e_j|)
\end{eqnarray}
which gives the following reduced density operator for $\ms + \ma $ :
\begin{eqnarray}
\rho^f_{\ms \ma} = Tr_{\me}(\rho^f_{\ms \ma \me}) = 
\sum_{j,k} c_j^* c_k (\sk \bsj)(\ak <a_j|) <e_j|e_k>.
\end{eqnarray}
Assuming [see Eq.(38)]
\begin{eqnarray}
<e_j|e_k> = \delta_{jk}
\end{eqnarray}
we have
\begin{eqnarray}
\rho^f_{\ms \ma}= \sum_j |c_j|^2 (\sj \bsj)(\aj <a_j|)
\end{eqnarray}
which is nothing but the density operator of Eq.(19) obtained from Eq.(18) 
by invoking the reduction process. Environment-induced decoherence, therefore, 
provides the dynamical mechanism for the reduction process.

\vspace{.15in}
\noindent
\textbf{4.4 Pointer basis of the quantum apparatus}

\vspace{.12in}
Coupling to the environment also solves the preferred basis problem 
described in section 3.3. This is because, in contrast to  the expansions of 
the form of Eq.(17) (which are generally non-unique), those of the form of 
Eq.(53) are unique.  This is guaranteed by the tridecompositional uniqueness 
theorem [25-27, 11] stated below.

\noindent
\textsl{ Tridecompositional uniqueness theorem :}\ If a vector $|\psi>$ in the 
Hilbert space $\mh_1 \otimes \mh_2 \otimes \mh_3 $ can be decomposed into the 
diagonal (``Schmidt'') form 
\begin{eqnarray}
|\psi> = \sum_i \alpha_i |\phi_i>_1 |\phi_i>_2 |\phi_i>_3 
\end{eqnarray}
the expansion is unique provided that $\{|phi_i>_1\}$ and  
$\{<\phi_i|_2\}$ are sets of linearly independent normalised vectors in 
$\mh_1$ and $\mh_2$ and that $\{|\phi_i>_3\}$ is a set of mutually 
noncollinear vectors in $\mh_3$.

This can be generalized to an N-decomposition uniqueness theorem where 
$N\geq3$.

This theorem, however, does not answer the question `Which basis is 
preferred ?' Zurek[2] provids the answer: The preferred pointer basis 
should be the basis (in \ha \ ) which contains a reliable record of the 
states of \ms \ , i.e. the basis $\{\aj\}$ for which the correlated states  
$\sj \aj$ are left undisturbed by the subsequent formation of correlations 
with the environment. 

To illustrate this, we consider pointer basis in a simple bit-by-bit 
measurement [3]. Both, the system \ms \ and the apparatus \ma \ are assumed 
to be two-state systems and a measurement interaction between them is 
assumed to give the correlated state which admits the two different 
expansions of Eq.(29a,b). Let us identify the apparatus \ma \ of this 
equation with the two-state system \msp \ of the example treated in section 
4.2 taking 
\begin{eqnarray}
|u> = |z,+>_{\ma} \hspace{.15in} and \hspace{.15in} |d> = |z,->_{\ma}.
\end{eqnarray}
Now, the interaction (40) between this apparatus and the environment \me \ 
(consisting of N two-state systems ) suppresses the interference terms 
between $|u>$ and $|d>$; these states (the `pointer states'), therefore, 
constitute a preferred basis (the `pointer basis'). Out of the two 
expansions in Eq.(29a,b), the one in (29a) represents the stable correlation 
between the system and apparatus and the measured observable is $F = 
\sigma_z$.

This decoherence-induced selection of the preferred set of pointer states 
that remain stable in the presence of the environment is called 
\emph{environment-induced superselection} or \emph{einselection}.

In the example above, there is an effective superselection operative in 
\ha \ (disallowing /suppressing superpositions of the states $|u>$ and 
$|d>$). Any operator of the form 
\begin{eqnarray}
G = \zeta_1 |u><u| + \zeta_2 |d><d|
\end{eqnarray}
(\emph{pointer observable}) with $\zeta_i$ real and distinct acts as a 
superselection operator. We present a systematic treatment of such 
\emph{environment-induced superselection rules} in the next subsection.

\vspace{.15in}
\noindent
\textbf{4.5 Environment-induced superselection rules}

\vspace{.12in}
We come back to the $\msp+\me$ system of section 4.2 and treat it more 
systematically [3]. The Hilbert space is $\mh = \mh_{\msp} \otimes 
\mh_{\me}$ and we choose  orthonormal bases $\{|n>\}$ in $\mh_{\msp}$ 
and $\{|e_j>\}$ in $\mh_{\me}$. The total Hamiltonian is taken as 
\begin{eqnarray}
H = \sum_n \delta_n |n><n| + \sum_j \epsilon_j |e_j><e_j| + H^{\msp \me}
\end{eqnarray}
where 
\begin{eqnarray*}
H^{\msp \me} = \sum_{n,j} \gamma_{nj} |n><n| \otimes |e_j><e_j|
\end{eqnarray*}
which has a diagonal form with eigenvalues $\gamma_{nj}$ associated with the 
eigenvectors $|n> |e_j>$. (For a justification for the omission of  
off-diagonal terms of the form  $|m><n| \otimes |e_j><e_k|$, see [3].) 

The initial state of the combined system is assumed to be a product state:
\begin{eqnarray}
|\Phi(t=0)> = |\phi_{\msp}> |\psi_{\me}> = (\sum_n \alpha_n |n> )
(\sum_j \beta_j |e_j>).
\end{eqnarray}
It evolves, at time t, to
\begin{eqnarray}
|\Phi(t)> = \sum_{n,j}\alpha_n \beta_j exp[-i(\delta_n + \epsilon_j + 
\gamma_{nj})t/\hbar] |n> |e_j>.
\end{eqnarray}
The corresponding reduced density operator for \msp \ is
\begin{eqnarray}
\rho^{\msp}(t) = Tr_{\me} (|\Phi(t)><\Phi(t)|) = \sum_{m,n} \rho_{mn}(t) 
|m><n|
\end{eqnarray}
where (putting $\hbar$ =1 )
\begin{eqnarray}
\rho_{mn}(t) = \alpha_m \alpha_n^* e^{-i(\delta_m - \delta_n)t} 
\sum_k |\beta_k|^2 e^{-i(\gamma_{mk}- \gamma_{nk})t}.
\end{eqnarray}
The diagonal terms are time-independent :
\begin{eqnarray}
\rho_{nn}(t) = |\alpha_n|^2 \sum_k|\beta_k|^2 = |\alpha_n|^2.
\end{eqnarray}
The off-diagonal terms have time dependence in the two exponentials. The 
first one is a trivial `rotation' (in the relevant complex plane); the more 
important is the second one contained in the `correlation amplitude' 
\begin{eqnarray}
z_{mn}(t) & = & \sum_k |\beta_k|^2 e^{-i(\gamma_{mk}-\gamma_{nk})t} \nonumber \\
& \equiv & \sum_k p_k e^{-i\omega^{mn}_k t}
\end{eqnarray}
where $ p_k = |\beta_k|^2$ and $\omega_k^{mn} = \gamma_{mk} -\gamma_{nk}$. 
Now, defining 
\begin{eqnarray}
<f(t)> = lim_{T \rightarrow \infty}\frac{1}{T}\int_{t}^{t+T}f(t^{\prime})
dt^{\prime}
\end{eqnarray}
we have 
\begin{eqnarray}
<z_{mn}(t)> = 0
\end{eqnarray}
and
\begin{eqnarray}
<|z_{mn}(t)|^2> = \sum_{k,k^{\prime}}p_k p_{k^{\prime}}
\delta (\omega_k^{mn}, \omega_{k^{\prime}}^{mn}).
\end{eqnarray}
Assuming, for simplicity, that all $\omega_k^{mn}$ are distinct and that 
there are N active states in the environment, we have the standard deviation 
$\Delta$ of the correlation amplitude given by 
\begin{eqnarray}
\Delta^2 = \sum_{k=1}^N p_k^2.
\end{eqnarray}
Assuming that all the $p_k$ are approximately equal (with $p_k \sim N^{-1}$), 
we have
\begin{eqnarray}
\delta \sim N^{-1/2}.
\end{eqnarray}
Eqs.(69) and (72) show that  large environments effectively damp out 
correlations between those states of the system which correspond to different 
eigenvalues of $H^{\msp \me}$. 

There may, in general, be more than one vectors in $\mh_{\msp}$ corresponding 
to the same eigenvalue $\gamma_{nj}$. These vectors span a subspace $\mh_r$ 
of $\mh_{\msp}$. We have 
\begin{eqnarray}
\mh_{\msp} = \oplus_r \mh_r.
\end{eqnarray}
Only superpositions of states in a single $\mh_r$ are protected under the 
environmental monitoring. (Superpositions of states from more than one 
$\mh_r$ would have off-diagonal terms in the density operator, which are 
not allowed.) This implies the operation of a superselection rule. 

A selfadjoint operator A on $\mh_{\msp}$ is an observable only if, acting on 
a physical state, it gives a physical state. This implies 
\begin{eqnarray}
|\phi> \in \mh_r \Rightarrow A|\phi> \in \mh_r  \hspace{.15in} for 
\hspace{.12in} all \hspace{.12in}r.
\end{eqnarray}
Observables which are (distinct) multiples of identity on the `coherent 
subspaces' $\mh_r$ are called superselection operators. They are of the form 
\begin{eqnarray}
\Lambda = \sum_r \zeta_r P_r
\end{eqnarray}
where $P_r$ is the projection operator on $\mh_r$ and $\zeta_r$ are (distinct) 
real numbers. These are the observables which serve to distinguish different 
coherent subspaces. If the system \msp \ is to serve as an apparatus, 
distinctness of different pointer positions will be protected under 
environmental monitoring if different pointer positions correspond to vectors 
in different coherent subspaces. Such states will be eigenstates of 
$\Lambda$ corresponding to distinct eigenvalues. For this reason, 
observables of the form (75)  are referred to as \emph{pointer observables} in 
the decoherence -related literature. A basis in $\mh_{\msp}$ which consists of 
a pointer observable (and, therefore, all pointer observables) is called a 
\emph{pointer basis}.

In a pointer basis, $H^{\msp \me}$ as well as all the $P_r$ are diagonal. It 
follows that
\begin{eqnarray}
[\Lambda, H^{\msp \me}] = 0.
\end{eqnarray}
[Being an operator equation, Eq.(76) is, of course, independent of the 
choice of basis.]

In practice, pointer observables relating to macroscopic apparatus are often 
position observables (whose eigenvalues correspond to pointer positions). 
This is related to the fact that typical interaction potentials are 
functions of position variables (and other operators like spin operators 
which commute with the position variables). For more details on this, see 
[3,6,7]. 

A question naturally arises whether the familiar superselection rules 
associated with electric charge and univalence [essentially $(-1)^{2J}$ 
where J = angular momentum] and possibly others could have their origin in 
the action of environment-induced decoherence at a deper level. For a 
detailed treatment of this topic, we refer to Giulini [28].

\vspace{.15in}
\noindent
\textbf{4.6 Decoherence in a soluble model; decoherence time scale [29,5]}

\vspace{.12in}
To illustrate some features of decoherence, we next consider a soluble model 
in one space dimension  in which the system is a harmonic oscillator with 
position coordinate q(t) and the environment is modelled as a scalar field 
$\phi(x,t)$. The action is
\begin{eqnarray}
I = \int \int dt dx \left\{ \frac{1}{2} \left[\dot{\phi}^2 - 
(\frac{\partial \phi}{\partial x})^2 \right] + \frac{1}{2} \delta(x) 
(m\dot{q}^2 - \Omega^2_0 q^2 - \epsilon q \dot{\phi}) \right\}.
\end{eqnarray}
The time derivative coupling between the field and the oscillator is taken to 
ensure simple damping behaviour for the oscillator in the coupled system.

At time t = 0, the oscillator and the field are assumed to be 
\mbox{uncorrelated:} 
\begin{eqnarray}
\rho_{OF}(0) = \rho_O(0) \rho_F(0).
\end{eqnarray}
We shall be concerned with the reduced density operator $\rho = 
Tr_F \rho_{OF}$ of the oscillator. In the calculation of (a Fourier 
transform of) the density matrix elements $\rho(q,q^{\prime},t)$, a 
logarithmic divergence appears which is tackled by introducing a high 
frequency cutoff $\Gamma$. The density operator for the field is taken to be 
the one corresponding to thermodynamic equilibrium at temperature T. In the 
high temperature limit
\begin{eqnarray}
T >> \Gamma >> max (\gamma, \Omega)
\end{eqnarray}
where
\begin{eqnarray}
\gamma = \frac{\epsilon^2}{4m}, \hspace{.3in} \Omega = 
\sqrt{\Omega_0^2 - \gamma^2}
\end{eqnarray}
the master equation for $\rho$ is 
\begin{eqnarray}
\frac{\partial}{\partial t} \rho (q, q^{\prime},t) = \left[ \frac{i}{\hbar} 
\left\{ \frac{\partial^2}{\partial q^2} - 
\frac{\partial^2}{\partial {q^{\prime}}^2} 
-\Omega^2(q^2 - {q^{\prime}}^2) \right\} - \right. \nonumber \\ 
  \left. \gamma (q-q^{\prime})
(\frac{\partial}{\partial q} -\frac{\partial}{\partial q^{\prime}}) - 
\frac{2m\gamma k_B T}{\hbar^2}(q-q^{\prime})^2 \right] 
\rho (q, q^{\prime},t).
\end{eqnarray}
In this equation, the term $ \{  \}$ is the von Neumann term $ -i\hbar^{-1} 
[H_0,\rho]$ (with the bare frequency $\Omega_0$ of the oscillator replaced by 
$\Omega$). The second term on the right causes dissipation. The last 
`quantum diffusion' term is the one that will be seen to be important in the 
context of decoherence.

Let us consider a Schr$\ddot{o}$dinger cat state for the oscillator given 
by the wave function 
\begin{eqnarray}
\psi(q) = \frac{1}{\sqrt 2} [\chi_+(q) + \chi_-(q)] 
\end{eqnarray}
where  
\begin{eqnarray*}
\chi_{\pm}(q) = <q|\chi_{\pm}> = 
A exp \left[- \frac{(q\pm \frac{\Delta q}{2})^2}{4\delta^2} \right].
\end{eqnarray*}
For wide separation between the peaks of the two Gaussian wave packets in the 
superposition (82) ($ \Delta x >> \delta$), the density matrix $\rho(q, 
q^{\prime}) = \psi(q) \psi(q^{\prime})^*$ has four peaks: two on the 
diagonal ($q = q^{\prime}$) and two off-diagonal. The presence of the latter 
two peaks signifies quantum coherence; decoherence will be seen as 
vanishing/decay of these peaks. This decay is caused by the last term in 
Eq.(81). Noting that, for the off-diagonal peaks, $ (q-q^{\prime})^2 \simeq 
(\Delta q)^2$, we have, for $\rho_{+-} = |\chi_+><\chi_-|$,
\begin{eqnarray}
\frac{d}{dt}\rho_{+-} \sim - \tau_D^{-1} \rho_{+-}
\end{eqnarray}
where 
\begin{eqnarray}
\tau_D = \tau_R [\frac{\lambda_{dB}}{\Delta q}]^2.
\end{eqnarray}
Here $\tau_R = \gamma^{-1}$ is the relaxation time and $\lambda_{dB} = 
\hbar (2mk_BT)^{-1/2}$ is the thermal de Broglie wave length.

For T= 300 K, m = 1gm, $\Delta q $ = 1 cm, we have 
\begin{eqnarray}
\frac{\tau_D}{\tau_R} \sim 10^{-40}.
\end{eqnarray}
Thus, even if the relaxation time $\tau_R$ were of the order of the age of 
the universe ($\sim 10^{17}$ sec), we have $\tau_D \sim 10^{-23}$ sec. For 
macroscopic systems, therefore, decoherence is an EXTREMELY FAST process. 
For microscopic systems (small m and $\Delta q$), $\tau_D$ is relatively 
large. It is also large at low temperatures.

\vspace{.15in}
\noindent
\textbf{4.7 Emergence of classicality [5]}

\vspace{.12in}
Classical behaviour of macroscopic systems that we normally come across 
must be explained in quantum mechanical terms becuse, as emphasized earlier, 
all systems in nature are quantum mechanical. (We live in a world with 
$\hbar \neq 0$.) For this, it is not adequate to show that, in a certain 
limit, appropriate quantum mechanical equations go over to the familiar 
classical equations. Discussion of classical behaviour of maroscopic systems 
involves states as well as equations of motion. A typical macroscopic object 
(a coin, for example) as a quantum mechanical system, occupies a small 
subset of the possible states in the quantum mechanical Hilbert space of the 
constuent particles. [In particular, Schr$\ddot{o}$dinger cat states 
(macroscopic superpositions) are absent.] We have seen how, in the context of a 
simple model, decoherence leads to effective elimination of 
Schr$\ddot{o}$dinger cat states. To show how the quantum mechanical 
equations of motion of such objects reduce to the classical ones, we shall 
consider the Wigner function of the oscillator of the previous section. 

Wigner transform of  a wave function $\psi(q)$ is defined as [30] 
\begin{eqnarray}
W_{\psi} (q,p) = \frac{1}{2\pi \hbar}\int_{-\infty}^{\infty}ds e^{ips/\hbar} 
\psi (q - \frac{s}{2}) \psi^*(q + \frac{s}{2}).
\end{eqnarray}
This object is real but not necessarily non-negative and therefore not a 
phase space probability density. We have, however, 
\begin{eqnarray} 
\int_{- \infty}^{\infty} W_{\psi}(q,p) dp = |\psi (q) |^2, \hspace{.15in} 
\int_{- \infty}^{\infty} W_{\psi} (q,p) dq = | \tilde{\psi}(p)|^2 
\end{eqnarray}
where $\tilde{\psi}$ is the Fourier transform of $\psi$. 

For the minimum uncertainty wave packet 
\begin{eqnarray}
\psi(q) = \pi^{-\frac{1}{4}} \delta^{- \frac{1}{2}} 
exp \left[ - \frac{(q - q_0)^2}
{2 \delta^2} + i p_0q \right]
\end{eqnarray}
$W_{\psi}$ is Gaussian in both q and p (and non-negative) :
\begin{eqnarray}
W_{\psi} (q,p) = \frac{1}{\pi \hbar}exp \left[ - \frac{(q - q_0)^2}{2 \delta^2} 
- \frac{(p - p_0)^2 \delta^2}{\hbar^2} \right]. 
\end{eqnarray}
It describes a system localised in both q and p. This is the closest 
approximation to a point in phase space that a wave function can yield. 

Note that, on the right hand side in Eq.(86), the dependence on $\psi$ is in 
terms of the corresponding density matrix. Generalizing this  to a general 
density matrix $\rho(q,q^{\prime})$, we have 
\begin{eqnarray}
W_{\rho} (q,p) = \frac{1}{2 \pi \hbar} \int_{- \infty}^{\infty} 
e^{ips/\hbar} \rho(q - \frac{s}{2}, q + \frac{s}{2}) ds.
\end{eqnarray}
Eq.(86) is a special case of (90) with $\rho(q,q^{\prime}) = 
\psi (q) \psi^* (q^{\prime})$. For 
\begin{eqnarray*}
\rho = \sum_{i = 1}^{n} p_i |\psi_i><\psi_i|
\end{eqnarray*}
where $p_i \geq 0$ and $|\psi_i>$ are minimum uncertainty wave packets, we have 
$W_{\rho} \geq 0$ and represents a probability density in phase space. 

Combining Eqs(81) and (89) to obtain a `master equation' for W (dropping the 
subscript $\rho$) and replacing the harmonic oscillator potential  by a 
general potential V(q), we have 
\begin{eqnarray}
\frac{\partial W(q,p,t)}{\partial t} = - \frac{p}{m} \frac{\partial W}
{\partial q} + \frac{\partial V_r(q)}{\partial q} \frac{\partial W}{\partial p}
 + 2 \gamma \frac{\partial}{\partial p}(pW) + 
 D \frac{\partial^2 W}{\partial p^2}
 \end{eqnarray}
 where $V_r$ (= V + extra terms) is the renormalized potential (recall the 
 replacement of $ \Omega_0$ by $\Omega$ in the previous subsection) and 
 $D = 2m \gamma k_B T$.

The first two terms on the right in Eq.(90) can be written as the Poisson 
bracket $\{H, W \}$. For large m, $\gamma$ is small. When the last term has done its main job (of suppressing the off-diagonal terms in the density matrix), 
its action on the  remaining (density operator)/(Wigner function) is negligible 
 and we have, finally, the desired classical equation
 \begin{eqnarray}
 \frac{\partial W}{\partial t} = \{ H, W \}.
 \end{eqnarray}

\vspace{.2in}
\noindent
\textbf{V. Does Decoherence Completely Solve the Measurement Problems ? }

\vspace{.12in}
The answer to the question posed above is closely tied up with the 
interpretation of quantum mechanics. 

By interpretation of the formalism of a scientific theory, one essentially 
means the explanation/translation of the new concepts and terms used in the 
theory in the commonly accepted logical framework (if necessary, by 
appropriately extending the existing logical framework). The need for 
interpretation in quantum mechanics arises mainly because it is an inherently 
probabilistic theory whose probabilistic aspects do not admit the traditional 
ignorance interpretation. Its kinematical framework employing novel objects 
like (state vectors)/(wave functions) and operator observables gives rise to 
questions like the interpretation of $\psi$ (whether it represents a single 
system or an ensemble), the question of assignment of definite values to 
observables, the relationship of the formalism with objective reality, etc. 

We shall be mainly concerned with the question: ` Is it possible to 
consistently interpret the formalism of quantum mechanics so that, taking 
into consideration decoherence-related developments, the measurement problems 
are completely solved ?'

Let us have a quick look at the relevant decoherence-related developments. 
Interaction of a (typically macroscopic) system with the environment selects 
a preferred basis (in the quantum mechanical Hilbert space of the system) 
consisting of a set of robust ( in the sense that they persist in the presence 
of continuous environmental monitoring) quasiclassical states characterised by 
eigenvalues of observables (approximately) commuting with the 
system-environment Hamiltonian. Since interactions are generally described in 
terms of position/configuration variables, the emerging quasiclassical 
properties generally involve localization of objects. The equations of motion 
for the class of quantum systems referred to above go over to the familiar 
deterministic classical equations.This serves to explain the appearance of 
determinate, objective (in the sense of the above-mentioned robustness) 
properties to a local observer.

One is now tempted to conclude that this (plus the specific results obtained 
in sections 4.3 and 4.4) must be adequate to explain, for all practical 
purposes, the observation of a unique stable pointer state at the end of a 
measurement. A critical look at the whole development ( see Bub [25] and Adler 
[33] for a careful discussion and detailed references) shows, however, that 
the answer to the question posed above must be in the negative. 

In section 4.3, the reduced density operator of system + apparatus is obtained 
by taking trace (over the enviroment) of the density operator of 
system + apparatus + environment. Zurek [5] interprets this as ignoring the 
uncontrolled and unmeasured degrees of freedom. This is supposed to be taken 
as similar to 
the procedure of deriving probability 1/2 for `heads' as well as for `tails' 
in the experiment of tossing a fair coin by averaging over the uncontrolled and unmeasured degrees of freedom of the environment of the coin. 

The two procedures are, in fact, substantially different [25]. In the coin 
toss experiment, when, ignoring the environment, we claim that the probability 
of getting `heads' in a particular toss of the coin is 1/2, we can also claim 
that we \emph{do} in fact get \emph{either} `heads' \emph{or} `tails' on each 
particular toss. A definite outcome can be predicted if we take into 
consideration appropriate enviromental parameters. 

In the case of a quantum measurement, however, we cannot claim that, taking 
the environment into consideration, a definite outcome of the experiment will 
be predicted. In fact, taking the environment into account will give us back 
the troublesome equation (53) from which we derived the mixed state (55) 
by tracing over the environment.

\vspace{.12in}
\textbf{What is/are the way(s) out ?}

\vspace{.12in}
\noindent 
\textbf{One route: (Relative state)/(Many worlds) interpretation of quantum 
mechanics (Everett, DeWitt and others [34])}

One insists that the superposition (17) \emph{is} the final outcome of the 
measurement. This equation is to be interpreted as a splitting of the state 
vector of (system + apparatus) into various branches (these are often 
called \emph{Everett branches}) only one of which we observe.

This approach is very uneconomical and intuitively unappealing. Moreover, 
the preferred basis problem is not solved in this approach.

\vspace{.12in}
\noindent
\textbf{Another route: Bohmian mechanics [35,36]}

In this approach, one has, apart from the wave function $\psi(q,t)$, the 
functions $q_{\alpha}(t)$ describing configuration space trajectories of the 
system. The wave function $\psi(q,t)$ serves as  a guidance field for 
the motion of the trajectories q(t) [essentially analogous to the way the 
Hamilton-Jacobi function S(q,t) serves, in classical mechanics, as a 
guidance field for the system trajectories in configuration space]. 

The functions $q_{\alpha}(t)$ serve as `hidden variables'. They serve to pick 
up unique outcomes in measurement situations. [At any given time, q(t) has a 
definite value. At the end of a measurement, the system trajectory is 
expected to be in any one of the various configuration domains corresponding 
to the different outcomes in the superposition (17).] Born rule probabilities 
emerge for the observer who cannot access the additional information contained 
in q(t). 

Problems with this approach : it fails in the relativistic domain and 
quantum field theory.

These two approaches are \emph{interpretations} of quantum mechanics in the 
specific sense that, by design, they reproduce all physical predictions of the 
traditional (nonrelativistic) quantum mechanics and so are empirically 
indistinguishable from the orthodox theory, while changing the formalism so as 
to resolve some difficulties of measurement theory.

If we insist on having one world existing within the standard arena of states 
and operators in Hilbert space, we must inject new physics in the formalism 
of quantum mechanics (and should be willing to discard, if necessary, one 
or more of the assumptions made in the traditional formalism.)

\vspace{.12in}
\noindent
\textbf{An alternative : Dynamical collapse models} (Ghirardi, Rimini and 
Weber [37], Pearle [38,39], Gisin [40] and Diosi [41]) 

In this approach, one abandons the assumption of a unitary evolution; this is 
replaced by a stochastic unitary one : 

\begin{eqnarray*}
d \psi (t) = (A dt + B dW_t) \psi (t) 
\end{eqnarray*}
where $W_t$ is a Wiener process and A and B are suitably chosen operators. 
Heuristically, the idea is that, quantum 
mechanics may be modified by a low level universal noise, akin to Brownian 
motion (possibly arising from physics at the Planck scale) which, in certain 
situations, causes reduction of the state vector. 

This approach reproduces the observed fact of discrete outcomes  governed 
by Born rule probabilities. It predicts the maintenance of coherence where 
it is observed (superconductive tunnelling, ...) while predicting state 
vector reduction in measurement situations [42,43]. For more details, the 
reader is referred to the literature cited above. 

\vspace{.2in}
\noindent
\textbf{ References} 

\vspace{.12in}
\begin{description}
\item[1] H.D. Zeh, Found. Phys. \textbf{1}, 69 (1970).
\item[2] W.H. Zurek, Phys. Rev. D \textbf{24}, 1516 (1981).
\item[3] W.H. Zurek, Phys. Rev. D \textbf{26}, 1862 (1982).
\item[4] E. Joos and H.D. Zeh, Z. Phys. B \textbf{59}, 223 (1985).
\item[5] W.H.Zurek, Phys. Today \textbf{44(10)}, 36 (1990); revised version : 
`Decoherence and the transition from quantum to classical-revisited', Los 
Alamos Science, Number 27, 2002 (arxiv : quant-ph/0306072).
\item[6] W.H. Zurek : `Preferred states, predictability, classicality and 
environment-induced decoherence',Prog. Theor. Phys. \textbf{89}, 281 (1993).
\item[7] W.H. Zurek : `Decoherence, einselection and  quantum origin of the 
classical', Rev. Mod. Phys. \textbf{75}, 715 (2003); 
(arxiv : quant-phy/0105127).
\item[8] D. Giulini et al (ed) : `Decoherence and the appearance of a 
classical world in quantum theory', Springer (1996); second edition [Joos 
et al (ed); 2003].
\item[9] R. Kaiser and C.West (ed) : `Coherent atomic matter waves', Les 
Houches  Lectures, Springer (2001).
\item[10] G. Bacciagaluppi: `The role of decoherence in quantum theory' in the 
Stanford Encyclopedia of Philisophy (Winter 2003 edition) ed. by E.N. Zalta; 
URL http://plato.stanford.edu/arxives/win2003/entries/qm-decoherence/.
\item[11] M. Schlosshauer : `Decoherence, the measurement problem and 
interpretations of quantum mechanics', arxiv : quant-ph/0312059.
\item[12] R. Omnes : `The Interpretation of Quantum Mechanics', Princeton 
University Press (1994). 
\item[13] R. Omnes : `Understanding Quantum Mechanics', Princeton University 
Press (1999). 
\item[14] J. von Neumann : `Mathematical Foundatins of Quantum Mechanics', 
Princeton University Press (1955).
\item[15] J. Kupsch :`Open Quantum Systems', in Ref.[8].
\item[16] J.A. Wheeler and W.H. Zurek : `Quantum Theory and Measurement',
Princeton University Press (1983).
\item[17] M. Jammer :`The Philosophy of Quantum Mechanics', Wiley, New York 
(1974).
\item[18] J.M. Jauch :`Foundations of Quantum Mechanics', Addison-Wesley (1968).
\item[19] F. London and E. Bauer :`La th$\acute{e}$orie d 'observationen 
M$\acute{e}$canique Quantique'(Herman, Paris); English translation in Wheeler 
and Zurek [16].
\item[20] E.P. Wigner : `The problem of measurement', Am. J. Phys. 
\textbf{31}, 6-15 (1963) (reprinted in [16]).
\item[21] E. Schr$\ddot{o}$dinger :`Die gegenw$\ddot{a}$rtige situation in 
der quantenmechanik', Naturwiss. \textbf{23}, 807-812, 823-828, 844-849 
(1935); English translation in [16]. 
\item[22] W.H. Zurek :`Preferrd sets of states, classicality and 
environment-induced decoherence' in Ref.[23].
\item[23] J.J. Halliwell, J. P$\acute{e}$rez-Mercader and W.H.zurek (ed): 
`Physical Origins of Time Asymmetry', Cambridge Univ. Press (1994). 
\item[24] E. Joos :`Decoherence through interaction with the environment' in 
Ref.[8].
\item[25] J. Bub : `Interpreting the Quantum World', Cambridge University 
Press (1997). 
\item[26] R. Clifton in `Symposium on the Foundations of Modern Physics 
1994-- 70 years of Matter Waves', Editions 
Frontiers, Paris , pp 45-60 (1995).
\item[27] A. Elby and J. Bub, Phys. Rev. A \textbf{49}, 4213 (1994).
\item[28] D. Giulini (with contribution from J. Kupsch) : `Superselection 
rules and symmetries' in [8].
\item[29] W.G. Unruh and W.H. Zurek, Phys. Rev. D \textbf{40}, 1071 (1989).
\item[30] E.P. Wigner, Phys. Rev. \textbf{40}, 749 (1932).
\item[31] B. d'Espagnat : `Conceptual Foundations of Quantum Mechanics', second edition, Benjamin (1976).
\item[32] A.A. Grib and W.A. Rodrigues Jr. : `Nonlocality in Quantum Physics', 
Kluwr Academic/Plenum Publishers, New York (1999).
\item[33] S.L. Adler: `Why decoherence has not solved the measurement problem: 
a response to P.W. Anderson' arxiv: quant-ph/0112095.
\item[34] B.S. Dewitt and N. Graham : `The Many Worlds Interpretation of 
Quantum Mechanics', Princeton Univ. Press (1973). 
\item[35] D. Bohm and B.J. Hiley: `The Undivided Universe: An Ontoogical 
Interpretation of Quantum Theory', Routledge, Chapman and Hall, London (1993). 
\item[36] S. Goldstein, Physics Today \textbf{51}, No. 3, 42 (1998); 
\textbf{51}, No. 4, 38 (1998).
\item[37] G.C. Ghirardi, A. Rimini and T. Weber, Phys. Rev. \textbf{D34},
 470 (1986).
\item[38] P. Pearle, Phys. Rev. \textbf{D13}, 857 (1976).
\item[39] P. Pearle, Int. J. Theor. Phys. \textbf{18}, 489 (1979).
\item[40] N. Gisin, Phys. Rev. Lett. \textbf{52}, 1657 (1984).
\item[41] L. Di$\acute{o}$si, Phys. Lett \textbf{A 129}, 419 (1988).
\item[42] S.L. Adler, J. Phys. A: Math. Gen \textbf{35}, 841 (2002).
\item[43] S.L. Adler:`Probability in orthodox quantum mechanics: probability 
as a postulate versus probability as an emergent phenomenon', arxiv: 
quant-ph/0004077 (2000).

\end{description}

\end{document}